\documentclass[aps,prb,twocolumn]{revtex4-2}
\usepackage[colorlinks=true, linkcolor=red, citecolor=blue, urlcolor=blue]{hyperref}
\usepackage{bm,latexsym,amssymb,amsmath,mathrsfs}
\usepackage{graphicx}
\usepackage{color}
\usepackage{times,setspace}

\newcommand{\up}{\uparrow}
\newcommand{\down}{\downarrow}
\newcommand{\tr}{{\rm tr}}
\newcommand{\bmr}{{\bm r}}
\newcommand{\bme}{{\bm e}}

\begin{document}

\title{Non-Hermitian Hubbard model without the sign problem}

\author{Tomoya Hayata}
\affiliation{Department of Physics, Keio University, Kanagawa, 223-8521, Japan}
\author{Arata Yamamoto}
\affiliation{Department of Physics, The University of Tokyo, Tokyo 113-0033, Japan}

\begin{abstract}
We study the Hubbard model with non-Hermitian asymmetric hopping terms.
The conjugate hopping terms are introduced for two spin components so that the negative sign is canceled out.
This ensures that the quantum Monte Carlo simulation is free from the negative sign problem.
We analyze the antiferromagnetic order and its suppression by the non-Hermiticity.
\end{abstract}

\maketitle

\section{Introduction}
Non-Hermitian quantum systems, that is, quantum systems whose Hamiltonian is no longer Hermitian due to interactions with environments or measurements, have been intensively discussed both theoretically and experimentally in recent years.
Quantum phenomena established in Hermitian quantum systems, such as topological phases of matter, have been extended to incorporate the non-Hermiticity (see e.g., Refs.~\cite{Ashida:2020dkc,RevModPhys.93.015005} for reviews).
Most of the previous studies concentrated on single-particle physics (including the excitations around mean fields), and the study of genuine many-body physics is still in the very early stages~\cite{PhysRevX.4.041001,2017NatCo...815791A,Lourenco:2018kvh,PhysRevLett.121.203001,PhysRevLett.123.123601}.
In particular, the development of reliable numerical methods to simulate them is in progress.

The quantum Monte Carlo methods are the most powerful numerical tools to study the nonperturbative properties of quantum many-body systems.
The expectation values of quantum operators are stochastically evaluated on the basis of the partition function.
The integrand of the partition function must be semi-positive definite, otherwise the negative sign problem harms the importance sampling.
Non-Hermiticity usually makes the partition function complex and violates the semi-positivity.
This is, however, not always the case.
For example, in the determinant quantum Monte Carlo algorithm~\cite{Blankenbecler:1981jt,DUANE1987216},  even if each fermion has a non-Hermitian matrix and complex-valued determinant, the complex phase can be canceled between multi-components of fermions and the total partition function can be semi-positive.
Such non-Hermitian quantum systems can be handled in the quantum Monte Carlo study.

In this paper, we study the Fermi-Hubbard model with non-Hermitian terms.
The Hubbard model is the simplest model to describe the magnetic properties of electrons in transition metals~\cite{doi:10.1098/rspa.1963.0204,PhysRevLett.10.159,10.1143/PTP.30.275}.
Despite its simplicity, it exhibits surprisingly rich phenomena such as the Mott insulator and antiferromagnetism, and it is also a model of high-temperature superconductors~\cite{RevModPhys.66.763}.
The conventional Hubbard model,  i.e., the one-band nearest-neighbor Hubbard model on a square lattice, is semi-positive at half filling, so that the quantum Monte Carlo simulation has achieved a great success~\cite{PhysRevB.31.4403,PhysRevB.40.506}.
We introduce asymmetric hopping terms, which are known in the Hatano-Nelson model~\cite{PhysRevLett.77.570,PhysRevB.56.8651}, to construct a non-Hermitian version of the Hubbard model \cite{PhysRevB.58.16051,PhysRevA.85.013610}.
Although the non-Hermitian Hubbard model is not semi-positive in general cases, our model is exceptionally free from the sign problem.

We focus on the effect of the asymmetric hopping terms on the antiferromagnetic order in the groundstate of the Hubbard model.
We utilize two theoretical analyses: the mean-field calculation and the quantum Monte Carlo simulation.
First, using the mean-field calculation, we reveal the mechanism that the non-Hermiticity suppresses the antiferromagnetic order.
Next, as full nonperturbative analysis, we compute the antiferromagnetic structure factor in the determinant quantum Monte Carlo simulation. 
The groundstate of the non-Hermitian Hubbard model can be realized by initially preparing the groundstate of the Hermitian Hubbard model and adiabatically ramping up the strength of the asymmetric hopping~\cite{2017NatCo...815791A}.  The method to observe (the suppression of) the antiferromagnetic order is the same as that in the Hermitian system. 

The rest of the paper is organized as follows. In Sec.~\ref{sec:model}, we describe our model and discuss the breakdown of the antiferromagnetism by the non-Hermiticity on the basis of the mean-field approximation. 
In Sec.~\ref{sec:QMC}, we show the result of the determinant quantum Monte Carlo simulation. 
Section~\ref{sec:Generalization} is devoted to discussion of more general conditions to get the sign-free non-Hermitian Hamiltonian.

\section{Non-Hermitian Hubbard model}
\label{sec:model}

We consider a non-Hermitian extension of the Fermi-Hubbard model; the hopping parameters are imbalanced and spin-dependent.
The model Hamiltonian is
\begin{equation}
\label{eqH}
\begin{split}
H =&-\sum_{\bmr,i} \left\{ \left(t+\kappa\delta_{ix}\right)c^\dagger_{\bmr} c_{\bmr+\bme_i}+\left(t-\kappa\delta_{ix}\right)c^\dagger_{\bmr+\bme_i} c_{\bmr}\right\} \\
&-\sum_{\bmr,i} \left\{\left(t-\kappa\delta_{ix}\right)d^\dagger_{\bmr} d_{\bmr+\bme_i}+\left(t+\kappa\delta_{ix}\right)d^\dagger_{\bmr+\bme_i} d_{\bmr}\right\} \\
&+\sum_{\bmr} U  \left(c^\dagger_{\bmr} c_{\bmr} -\frac12\right) \left(d^\dagger_{\bmr} d_{\bmr}-\frac12\right) .
\end{split}
\end{equation}
There are three parameters: $t$ is the symmetric hopping parameter, $\kappa$ is the asymmetric hopping parameter, and $U$ is the coupling constant of the on-site repulsion.
The asymmetric hopping term breaks the Hermiticity of the Hamiltonian.
In this paper, we consider the two-dimensional square lattice $\bmr = (x,y)$ and the imbalanced hopping only in the $x$ direction, while these are generalizable to higher or lower dimensions.

By the continuous Hubbard-Stratonovich transformation, the Hamiltonian is written as
\begin{equation}
 H = H_\uparrow + H_\downarrow + \sum_{\bmr} \frac{U}{2} s^2_{\bmr} ,
\end{equation}
with
\begin{equation}
\begin{split}
\label{eqHup}
 H_\uparrow&=-\sum_{\bmr,i} \bigg\{ \left(t+\kappa\delta_{ix}\right)c^\dagger_{\bmr} c_{\bmr+\bme_i}+\left(t-\kappa\delta_{ix}\right)c^\dagger_{\bmr+\bme_i} c_{\bmr} \\
&\quad +Us_{\bmr}\left(c^\dagger_{\bmr} c_{\bmr} -\frac12\right)\bigg\} ,
\end{split} 
\end{equation}
and
\begin{equation}
\begin{split}
\label{eqHdown1}
 H_\downarrow&=-\sum_{\bmr,i} \bigg\{ \left(t-\kappa\delta_{ix}\right)d^\dagger_{\bmr} d_{\bmr+\bme_i}+\left(t+\kappa\delta_{ix}\right)d^\dagger_{\bmr+\bme_i} d_{\bmr} \\
&\quad -Us_{\bmr}\left(d^\dagger_{\bmr} d_{\bmr}-\frac12\right)\bigg\},
\end{split}
\end{equation}
where $s_{\bmr}$ is the auxiliary field.
The Hamiltonians \eqref{eqHup} and \eqref{eqHdown1} are not Hermitian at nonzero $\kappa$, so their energy spectra are complex.

\begin{figure}[t]
\begin{center}
 \includegraphics[width=.48\textwidth]{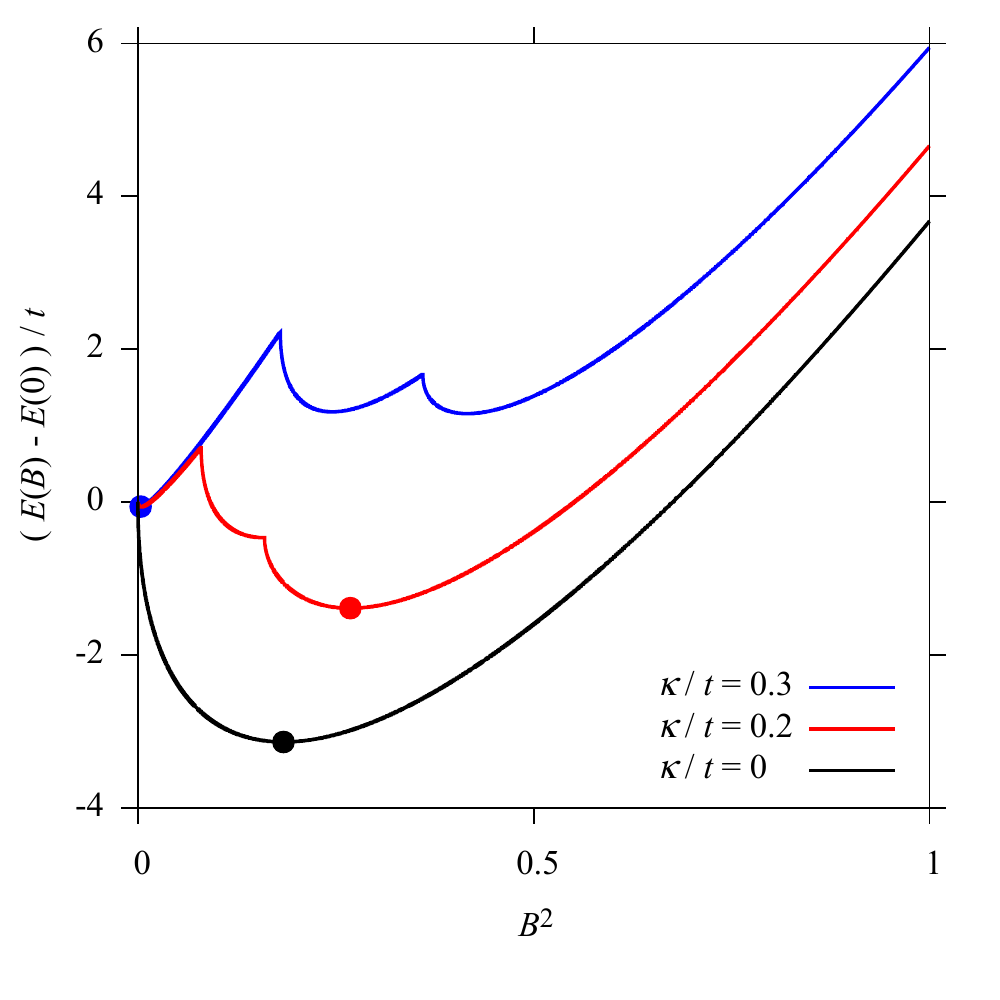}
\caption{\label{figEB}
Ground-state energy $E$ as a function of the antiferromagnetic mean field $B$.
The coupling constant is $U/t=1$ and the lattice volume is $V=8^2$.
The solid circles are the minima.
}
\end{center}
\end{figure}
\begin{figure}[h]
\begin{center}
 \includegraphics[width=.48\textwidth]{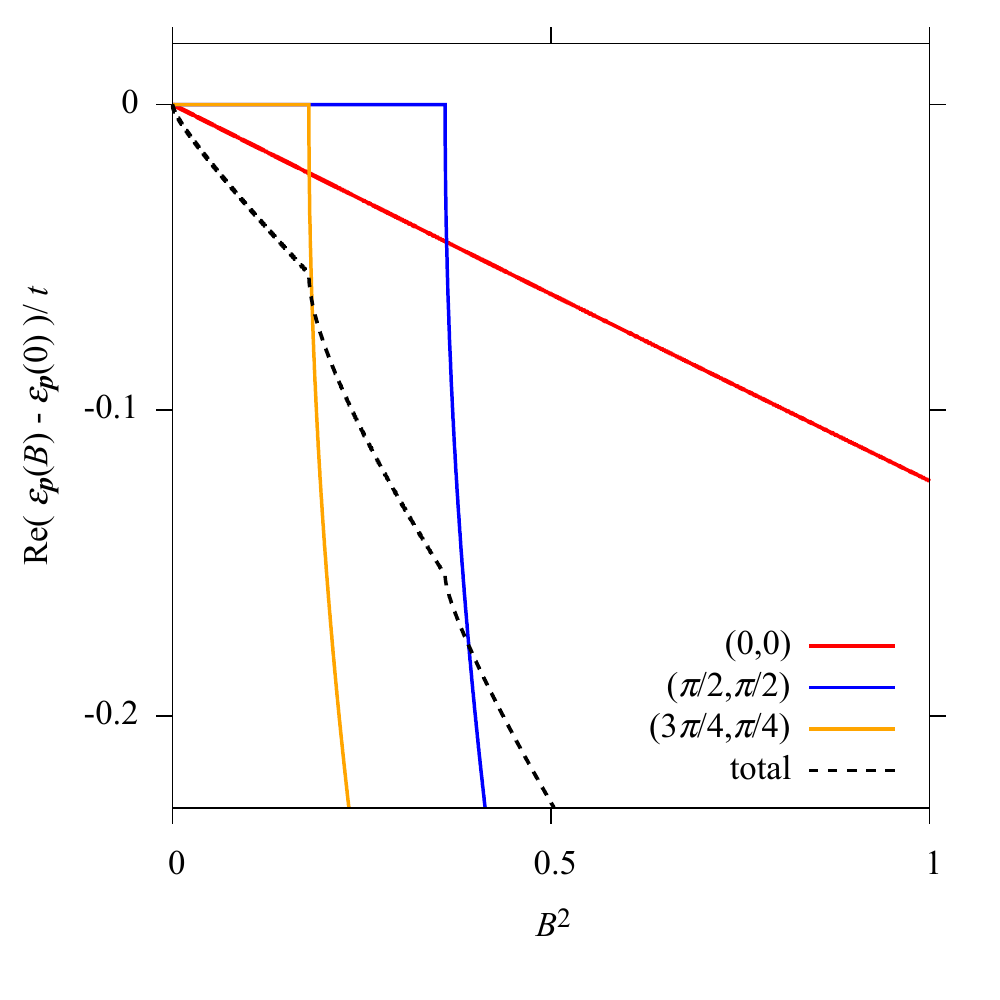}
\caption{\label{figEBP}
Single-particle fermion energy $\varepsilon_{\bm p}$ with the momentum $(p_x,p_y)$. 
The ``total'' stands for $2{\rm Re} \sum_{\bm p} (\varepsilon_{\bm p}(B)-\varepsilon_{\bm p}(0)) / V$.
The parameters are $U/t=1$, $\kappa/t=0.3$, and $V=8^2$.
}
\end{center}
\end{figure}

The repulsive Hubbard model exhibits antiferromagnetism.
The generation of the antiferromagnetic order can be analyzed in the mean-field approximation \cite{RevModPhys.66.763}.
The mean field is defined by the staggered magnetization
\begin{equation}
B=\left\langle (-)^{x+y}(c^\dagger_{\bmr} c_{\bmr} - d^\dagger_{\bmr} d_{\bmr})\right\rangle. 
\end{equation}
The single-particle fermion energies are given by
\begin{equation}
 \varepsilon_{\uparrow \bm p} = \pm \sqrt{ 4\left\{t(\cos p_x + \cos p_y) + i\kappa \sin p_x \right\}^2 + U^2B^2 } ,
\end{equation}
and
\begin{equation}
 \varepsilon_{\downarrow \bm p} = \pm \sqrt{ 4\left\{t(\cos p_x + \cos p_y) - i\kappa \sin p_x \right\}^2 + U^2B^2 } .
\end{equation}
The positive and negative signs are taken for the excited and ground states, respectively. 
The dispersion relation of each spin is complex, but the sum $\varepsilon_{\uparrow \bm p}+\varepsilon_{\downarrow \bm p}$ is real.
The ground state has the total energy
\begin{equation}
\label{eqE}
 E = 2 {\rm Re} \sum_{\bm p} \varepsilon_{\bm p} + \frac{U}{2} B^2 V
\end{equation}
with 
\begin{equation}
 \varepsilon_{\bm p} = - \sqrt{ 4\left\{t(\cos p_x + \cos p_y) + i\kappa \sin p_x \right\}^2 + U^2B^2 }  .
\end{equation}
The momentum sum $\sum_{\bm p}$ is performed only in half of the Brillouin zone to satisfy half filling.
The ground-state energy is shown in Fig.~\ref{figEB}.
At $\kappa=0$, these equations reproduce the famous properties of the Hermitian Hubbard model; the antiferromagnetic order $B\neq 0$ is energetically favored and the fermion spectrum is gapped.
When the non-Hermitian parameter $\kappa$ is turned on, the solution with $B=0$ appears.
The solution with $B\neq 0$ still exists but becomes metastable.
A discontinuous jump from $B\neq 0$ to $B= 0$, i.e., the first-order phase transition, happens at a certain value of $\kappa$.
Another curious behavior is the appearance of cusps in the dispersion relations.
These behaviors originate from the non-Hermiticity.
Let us consider the fermion modes with $\cos p_x + \cos p_y=0$.
The dispersion relation
\begin{equation}
 \varepsilon_{\bm p} =
 \begin{cases}
  - i \sqrt{ 4\kappa^2 \sin^2 p_x - U^2B^2} & (B^2 < B_{\rm ex}^2)\\
  - \sqrt{ - 4\kappa^2 \sin^2 p_x + U^2B^2} & (B^2 > B_{\rm ex}^2)
 \end{cases}
\end{equation}
turns from pure imaginary to real at the exceptional point $B_{\rm ex}^2 \equiv 4\kappa^2 \sin^2 p_x / U^2$.
Typical cases are shown in Fig.~\ref{figEBP}.
The mode of the momentum $(p_x,p_y)=(3\pi/4,\pi/4)$ makes the cusp at $B_{\rm ex}^2 = 2\kappa^2 / U^2 = 0.18t$ (yellow line) and the mode of $(\pi/2,\pi/2)$ makes the cusp at $B_{\rm ex}^2 = 4\kappa^2 / U^2 = 0.36t$ (blue line).
Since the imaginary part does not contribute to the total energy \eqref{eqE}, these modes do not favor the nonzero value of $B$ below the exceptional point.
Therefore the antiferromagnetic order cannot be formed at large $\kappa$.

The mean-field analysis predicts that the non-Hermitian term suppresses the antiferromagnetic order.
The prediction should be verified in full quantum analysis.
We do that by the quantum Monte Carlo method in the next section.

\section{Quantum Monte Carlo}
\label{sec:QMC}

We performed the determinant quantum Monte Carlo simulation \cite{Blankenbecler:1981jt}.
The partition function $Z = {\rm Tr} \left( e^{-\beta H} \right)$ is written in the imaginary-time formalism.
By the Suzuki-Trotter decomposition and the discrete Hubbard-Stratonovich transformation \cite{PhysRevB.40.506}, it becomes
\begin{equation}
\label{eqZHS}
\begin{split}
 Z=&\sum_{\{s_{\bmr}(\tau)\}}\tr\prod_{\tau}e^{-\Delta\tau K_\up}e^{- V_\up(\tau)}\tr\prod_{\tau}e^{-\Delta\tau K_\down}e^{-V_\down(\tau)}.
\end{split}
\end{equation}
The auxiliary field takes two values $s_{\bmr}(\tau)=\pm 1$ and the imaginary time is discretized as $\tau=0,\Delta\tau,\cdots,\beta-\Delta\tau$.
The spin Hamiltonians are given by
\begin{eqnarray}
 K_\uparrow & =& -\sum_{\bmr,i} \left(t+\kappa\delta_{ix}\right)c^\dagger_{\bmr} c_{\bmr+\bme_i}+\left(t-\kappa\delta_{ix}\right)c^\dagger_{\bmr+\bme_i} c_{\bmr}
\\
K_\downarrow &=& -\sum_{\bmr,i}\left(t-\kappa\delta_{ix}\right)d^\dagger_{\bmr} d_{\bmr+\bme_i}+\left(t+\kappa\delta_{ix}\right)d^\dagger_{\bmr+\bme_i} d_{\bmr}
\end{eqnarray}
and
\begin{eqnarray}
 V_\uparrow(\tau) &=&\sum_{\bmr} \lambda s_{\bmr}(\tau) \left( c^\dagger_{\bmr} c_{\bmr} - \frac12 \right)
\\
 V_\downarrow(\tau)&=&-\sum_{\bmr} \lambda s_{\bmr}(\tau) \left( d^\dagger_{\bmr} d_{\bmr} - \frac12 \right)
\end{eqnarray}
where $\cosh\lambda=e^{U\Delta\tau/2}$.
On a bipartite lattice and at half-filling, the spin-down Hamiltonian can be rewritten as
\begin{equation}
 K_\downarrow = -\sum_{\bmr,i}\left(t+\kappa\delta_{ix}\right)d^\dagger_{\bmr} d_{\bmr+\bme_i}+\left(t-\kappa\delta_{ix}\right)d^\dagger_{\bmr+\bme_i} d_{\bmr}
\end{equation}
and
\begin{equation}
 V_\downarrow(\tau)=\sum_{\bmr} \lambda s_{\bmr}(\tau) \left( d^\dagger_{\bmr} d_{\bmr} - \frac12 \right)
\end{equation}
by the particle-hole transformation $d^\dagger_{\bmr} \leftrightarrow (-1)^{x+y} d_{\bmr}$.
This is the same as the spin-up Hamiltonian.
Therefore, the partition function is equivalent to
\begin{equation}
\begin{split}
 Z  &= \sum_{\{s_{\bmr}(\tau)\}} \left[ \tr\prod_{\tau}e^{-\Delta\tau K_\up}e^{- V_\up(\tau)} \right]^2 \\
 &= \sum_{\{s_{\bmr}(\tau)\}}e^{\sum_{\tau,\bmr} \lambda s_{\bmr}(\tau)} \left[ \det\left(1+\prod_{\tau}e^{-\Delta\tau k_\up}e^{- v_\up(\tau)}\right) \right]^2 ,
\end{split}
\end{equation}
where $k_\up$ and $v_\up$ are defined as $K_\up=\sum_{{\bmr},{\bmr'}}c^\dagger_{\bmr} [k_\up]_{{\bmr}{\bmr'}}c_{\bmr'}$ and $V_\up(\tau)=\sum_{{\bmr},{\bmr'}}c^\dagger_{\bmr} [v_\up(\tau)]_{{\bmr}{\bmr'}}c_{\bmr'}-\sum_{\bmr}\lambda s_{\bmr}(\tau)/2$, respectively.
The squared form ensures semi-positivity because each determinant is real.

We adopted the stabilization techniques to keep the numerical accuracy during the computation of a long matrix chain associated with the imaginary-time evolution~\cite{PhysRevB.40.506,BAI2011659}.
We fixed the Suzuki-Trotter discretization with $\Delta\tau=0.1/t$.
The discretization errors are small enough compared with the statistical errors in the final results.
The statistical errors were estimated by the jackknife method.

\begin{figure}[t]
\begin{center}
 \includegraphics[width=.48\textwidth]{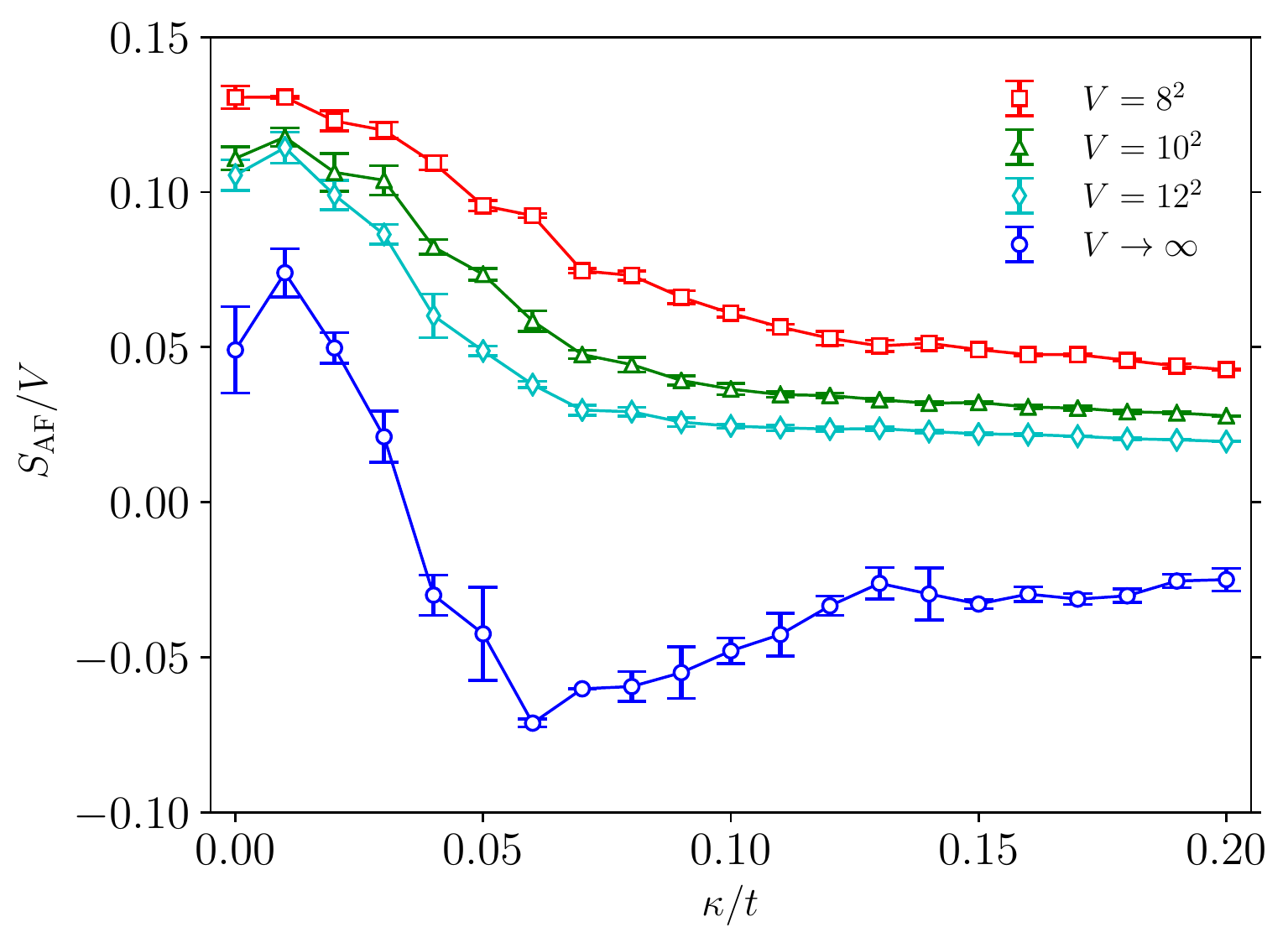}
\caption{\label{figQMC}
Antiferromagnetic structure factor $S_{\rm AF}/V$ as a function of $\kappa$.
The positive value of $S_{\rm AF}/V$ in the thermodynamic limit $V\to\infty$ implies the presence of the antiferromagnetic order.
The coupling constant is $U/t=4$.
The error bars are the statistical errors.
}
\end{center}
\end{figure}
\begin{figure}[h]
\begin{center}
 \includegraphics[width=.48\textwidth]{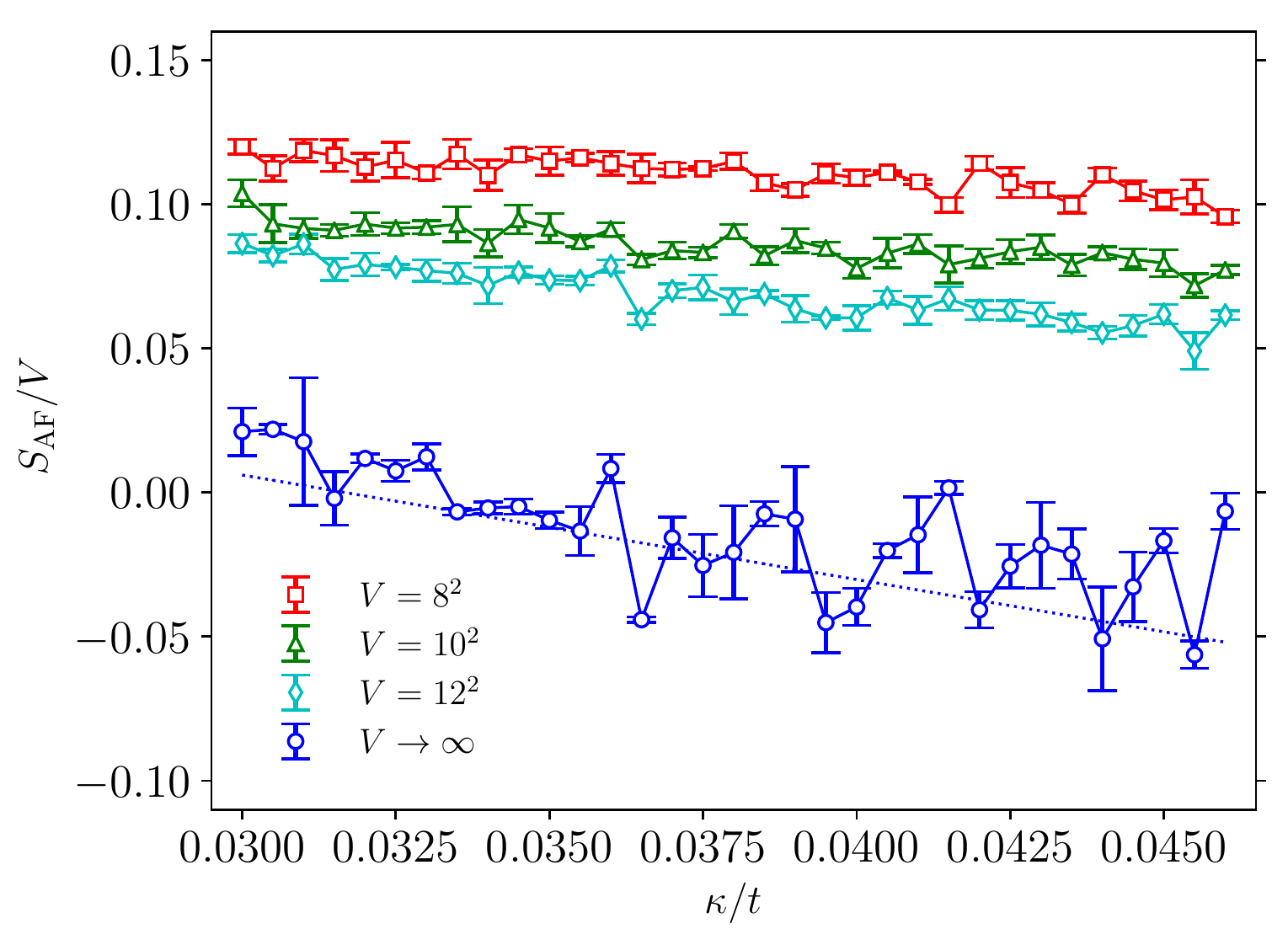}
\caption{\label{figQMC_critical}
Antiferromagnetic structure factor $S_{\rm AF}/V$ near the critical $\kappa$.
The blue dotted-line shows the weighted least-squares linear fit of the thermodynamic limit $V\rightarrow\infty$.
The coupling constant is $U/t=4$.
The error bars are the statistical errors.
}
\end{center}
\end{figure}
We computed the antiferromagnetic structure factor of the ground state
\begin{equation}
\frac{S_{\rm AF}}{V}=\lim_{\beta\to\infty}\left\langle \left[\frac{1}{V}\sum_{\bmr}(-)^{x+y}(c^\dagger_{\bmr} c_{\bmr} - d^\dagger_{\bmr} d_{\bmr})\right]^2\right\rangle .
\end{equation}
In the mean-field approximation, the structure factor is given by $S_{\rm AF}/V=B^2$.
We performed the simulation at $\beta t=12$, $14$, $16$, and $20$, and then extrapolated to the zero temperature $\beta t\rightarrow \infty$.
The obtained results are shown in Fig.~\ref{figQMC}.
The data with three lattice volumes $V=8^2$, $10^2$,  and $12^2$ are plotted.
To see whether the antiferromagnetic order survives in the thermodynamic limit, the data were extrapolated to $V\to\infty$ by the linear least-squares fitting of the $1/\sqrt{V}$ dependence of $S_{\rm AF}/V$ as $S_{\rm AF}/V=a+b/\sqrt{V}$according to the spin-wave theory~\cite{PhysRevB.40.506}.
The data point at $\kappa=0$ is consistent with the known result $\lim_{V\to\infty}S_{\rm AF}/V \simeq 0.05$ ~\cite{PhysRevB.40.506}.
The presence of the antiferromagnetic order can be judged by the sign of $\lim_{V\to\infty}S_{\rm AF}/V$; if it is positive, the antiferromagnetic order exists, and if it is negative, the antiferromagnetic order does not exist~\cite{PhysRevB.40.506}.
We can see that the antiferromagnetic order disappears with increasing $\kappa$.
The region near the critical $\kappa$ is magnified in Fig.~\ref{figQMC_critical}.
We estimated the critical $\kappa$ as $\kappa_c=0.0317\pm0.0014$ by the linear fit near the transition point, although the data largely fluctuate.
We cannot identify the reason for the fluctuation but it could be due to statistical and systematic errors in the double extrapolation.

We obtained qualitatively consistent results with the mean-field prediction.
Non-Hermiticity destroys the antiferromagnetic order.
At the quantitative level, the critical value of $\kappa$ is largely different between the mean-field and quantum Monte Carlo calculations.
This is not very surprising because two dimensions are the lowest bound for the phase transition and the opposite limit to the validity of the mean-field approximation.
Our results suggest that the antiferromagnetic order is more easily destroyed in the full quantum analysis than in the mean-field analysis.

\section{Generalization}
\label{sec:Generalization}
In this paper, we studied the non-Hermitian Hubbard model without the sign problem.
The model can be extended to other forms of the Hamiltonian.
The total fermion determinant is semi-positive definite as long as two fermion determinants are complex conjugate.
The conjugate pair is generalizable to
\begin{equation}
\begin{split}
 H_\uparrow&=\sum_{\bmr,i} \bigg\{ A_{\bmr,i}c^\dagger_{\bmr} c_{\bmr+\bme_i}+B_{\bmr,i}c^\dagger_{\bmr+\bme_i} c_{\bmr} \\
&\quad +C_{\bmr}\left(c^\dagger_{\bmr} c_{\bmr} -\frac12\right)\bigg\}
\end{split} 
\end{equation}
and
\begin{equation}
\begin{split}
 H_\downarrow&=\sum_{\bmr,i} \bigg\{ B^*_{\bmr,i}d^\dagger_{\bmr} d_{\bmr+\bme_i}+A^*_{\bmr,i}d^\dagger_{\bmr+\bme_i} d_{\bmr} \\
&\quad +C^*_{\bmr}\left(d^\dagger_{\bmr} d_{\bmr}-\frac12\right)\bigg\}.
\end{split}
\end{equation}
The coefficients $A_{\bmr,i}$, $B_{\bmr,i}$ and $C_{\bmr,i}$ are complex.
This general form, for example, includes the current-current interaction; the Hubbard-Stratonovich transformation is given by
\begin{equation}
 e^{gJ_{\up\bmr,i}J_{\down\bmr,i}} = \int dn \ e^{-\frac{g}{2} n^2 - g n(J_{\up\bmr,i}+J_{\down\bmr,i})}
\end{equation}
with
\begin{eqnarray}
 J_{\up\bmr,i} &=& i(c^\dagger_{\bmr} c_{\bmr+\bme_i}-c^\dagger_{\bmr+\bme_i} c_{\bmr}) ,\\
 J_{\down\bmr,i} &=& i(d^\dagger_{\bmr} d_{\bmr+\bme_i}-d^\dagger_{\bmr+\bme_i} d_{\bmr}).
\end{eqnarray}
Combining two conjugate fermions is a standard strategy to construct sign-problem-free fermion systems.
This standard strategy is well known and a more general strategy is also known in lattice quantum chromodynamics \cite{Hayata:2017jdh}.
They might be useful for finding other classes of sign-problem-free non-Hermitian systems.

\begin{acknowledgments}
This work was supported by JSPS KAKENHI Grant Numbers 19K03841, 21H01007, and 21H01084.
The numerical calculations were carried out on Yukawa-21 at YITP, Kyoto University, and on cluster computers at iTHEMS, RIKEN.
\end{acknowledgments}

\bibliographystyle{apsrev4-2}
\bibliography{paper}

\end{document}